\documentclass[aps,twocolumn,nofootinbib,superscriptaddress,preprintnumbers,prd]{revtex4-2}

\usepackage{amsmath,amssymb}
\usepackage{accents}
\usepackage{nccmath}
\usepackage{float}
\usepackage{color}
\usepackage[bookmarks=true,hyperfigures=true]{hyperref}
\usepackage{graphicx}
\usepackage[export]{adjustbox}
\usepackage{tensor}
\usepackage{empheq}
\usepackage{nicefrac}
\usepackage{shuffle}
\usepackage{slashed}

\allowdisplaybreaks[3]

\usepackage[font=small, labelfont=bf]{caption}

\begin{document}
\newtheorem{theorem}{Theorem}
\newtheorem{theorem1}{Theorem}
\thispagestyle{empty}


\title{Geodesics From Classical Double Copy}
\author{Riccardo Gonzo}
\affiliation{School of Mathematics \& Hamilton Mathematics Institute,
    Trinity College Dublin, College Green, Dublin 2, Ireland}
\author{Canxin Shi}
\affiliation{Institut f\"ur Physik und IRIS Adlershof, Humboldt-Universit\"at zu Berlin,
    Zum Gro{\ss}en Windkanal 2, 12489 Berlin, Germany}
\email{gonzo@maths.tcd.ie, canxin@physik.hu-berlin.de}
\preprint{SAGEX-21-18-E,
    HU-EP-21/25}

\begin{abstract}
We extend the Kerr-Schild double copy to the case of a probe particle moving in the Kerr-Schild background.
In particular, we solve Wong's equations for a test color charge in a Coulomb non-Abelian potential ($\sqrt{\text{Schw}}$) and on the equatorial plane for the potential generated by a rotating disk of charge known as the single copy of the $\text{Kerr}$ background ($\sqrt{\text{Kerr}}$).
The orbits, as the corresponding geodesics on the gravity side, feature elliptic, circular, hyperbolic and plunge behaviour for the charged particle.
We then find a new double copy map between the conserved charges on the gauge theory side and the gravity side,
which enables us to fully recover geodesic equations for Schwarzschild and Kerr. Interestingly, the map works naturally for both bound and unbound orbits.
\end{abstract}

\maketitle

\section{Introduction}

The double copy relation between gauge and gravity theories was first discovered for quantum scattering amplitudes \cite{Kawai:1985xq,Bern:2008qj,Bern:2010ue} and in recent years it quickly became a formidable tool to tame the complexity of perturbative gravity calculations \cite{Bern:2019prr}.
While initially applications of color-kinematics duality were based on the BCJ relation, there have been many further developments in understanding the symmetry principles \cite{Monteiro:2011pc,Bjerrum-Bohr:2012kaa,Tolotti:2013caa,Monteiro:2013rya,Cheung:2016prv,Anastasiou:2014qba,Anastasiou:2017nsz,Anastasiou:2016csv,Anastasiou:2018rdx,Mizera:2019blq,Borsten:2020zgj,Ferrero:2020vww,Borsten:2021hua,Chen:2021chy,Beneke:2021ilf,Cheung:2021zvb,Borsten:2021rmh} and in the implementation of the perturbative double copy in new contexts like the worldline formalism \cite{Goldberger:2016iau,Goldberger:2017frp,Goldberger:2017ogt,Shen:2018ebu,Plefka:2018dpa,Plefka:2019hmz,Bastianelli:2021rbt}, celestial amplitudes \cite{Casali:2020vuy,Pasterski:2020pdk} and perturbation theory on special backgrounds \cite{Adamo:2017nia,Adamo:2018mpq,Adamo:2020qru}.

Perhaps surprisingly, it turns out that also some exact classical solutions of Einstein equations can be obtained from gauge theory solutions in Yang-Mills (YM) theory.
One prominent idea is the Kerr-Schild double copy, first discovered by Monteiro, O'Connell and White \cite{Monteiro:2014cda} and later developed by a number of authors \cite{Kim:2019jwm,Luna:2015paa,Luna:2016due,Luna:2016hge,Ridgway:2015fdl,White:2016jzc,Carrillo-Gonzalez:2017iyj,DeSmet:2017rve,Ilderton:2018lsf,Berman:2018hwd,Luna:2018dpt,CarrilloGonzalez:2019gof,Bah:2019sda,Alawadhi:2019urr,Elor:2020nqe,Keeler:2020rcv,Easson:2020esh,Alawadhi:2020jrv,Monteiro:2020plf,Alkac:2021bav,Chacon:2021hfe}.
It was also clarified in  \cite{Luna:2018dpt} that the Kerr-Schild double copy is actually a special case of the Weyl double copy, which was recently proven using twistorial techniques \cite{White:2020sfn,Chacon:2021wbr,Chacon:2021hfe}.
The simplest example is provided by the map between the Coulomb-like YM solution (i.e. the non-Abelian $\frac{1}{r}$ potential) to the Schwarzschild spacetime.
The single-copy version of the Kerr metric can also be found, and it corresponds to the potential field generated by a rotating disk of color charge as explained in \cite{Monteiro:2014cda} (see also Appendix \ref{se:AppendixA}). We denote those solutions as $\sqrt{\text{Schw}}$ and $\sqrt{\text{Kerr}}$, respectively.

Following the discovery of gravitational waves, there was a renewed interest in getting classical observables from on-shell scattering amplitudes. In particular, a double copy map has been explored for the impulse and the spin kick for probe particles in the Kerr background \cite{Arkani-Hamed:2019ymq}. A striking connection of those observables with minimally coupled three-point amplitudes of massive particles \cite{Arkani-Hamed:2017jhn} with large (classical) spin has been noticed in \cite{Guevara:2018wpp,Guevara:2019fsj,Arkani-Hamed:2019ymq} and further developed in \cite{Chung:2018kqs,Moynihan:2019bor,Emond:2020lwi,Aoude:2020mlg,delaCruz:2020bbn,Cristofoli:2020hnk,Guevara:2020xjx,Bautista:2021wfy,Chiodaroli:2021eug,Aoude:2021oqj,delaCruz:2021gjp}. The double copy has now become also a useful tool in the calculation of observables of gravitational interest for the binary problem, in particular for the construction of integrands \cite{Luna:2017dtq,Kosower:2018adc,Maybee:2019jus,Johansson:2019dnu,Bautista:2019evw,Bern:2019crd,Bern:2019nnu,PV:2019uuv,Plefka:2019wyg,Carrasco:2020ywq, Carrasco:2021bmu,Cheung:2020gbf,Haddad:2020tvs,Herrmann:2021lqe,Herrmann:2021tct,Bern:2021dqo,Brandhuber:2021kpo,Brandhuber:2021eyq}. While the scattering problem naturally maps to hyperbolic orbits, there is an interesting analytic continuation \cite{Kalin:2019rwq,Kalin:2019inp,Kalin:2020mvi} which makes it possible to directly derive results for the corresponding bound cases \cite{Dlapa:2021npj}.

The classical YM theory is usually studied as a toy model for gravity, but it is also important by itself. One example is provided by the equations of motion of classical YM theory that describe the dynamics of the quark-gluon plasma, which is believed to be the predominant phase of matter before the entire universe was formed \cite{Heinz:1983nx, Litim:2001db, Elze:1989un}. In particular, for the description of high energy heavy ion collisions, the gluon field is also treated classically as a first approximation \cite{Iancu:2003xm, McLerran:1993ka, McLerran:1993ni, Gelis:2021zmx}.

In this work, we exploit the Kerr-Schild double copy to understand the relationship between the geodesic equations for a probe particle in the gravity background with a test charged particle moving in the corresponding gauge background. In detail, we solve Wong's equations \cite{Wong:1970fu} exactly for the $\sqrt{\text{Schw}}$ background and for equatorial orbits in $\sqrt{\text{Kerr}}$ by identifying the relevant conserved charges in terms of which the final solution can be expressed. Interestingly, we find a direct map of the conserved charges for probe particles in Schwarzschild and Kerr which makes possible to recover geodesic equations.

\section{Double copy of the conserved charges}
In this section we derive the relation between the conserved charges of a test charged particle in the YM potential and the corresponding charges for a probe particle in the Kerr-Schild gravitational background. For convenience, we express the non-Abelian gauge potential in a suitable basis of generators of the $\text{SU}(N)$ algebra
\begin{align}
\mathbb{A}_{\mu}= A_{\mu}^a \mathbb{T}^a.
\end{align}
A point charge moving in a YM background $A_\mu^a(x)$ is governed by the worldline Lagrangian \cite{Balachandran:1976ya,Balachandran:1977ub}
\begin{align} \label{eq:LagrangianYM}
    L^{\mathrm{YM}} = \frac{\bar{g}_{\mu\nu} v^\mu v^\nu }{2e} - \frac{e m^2}{2}
    + i \psi^\dagger \frac{d\psi}{d\tau}
    - g\, c^a v^\mu A_\mu^a(x),
\end{align}
where $v^\mu = dx^\mu/d\tau$, $e(\tau)$ is the einbein, $\bar{g}_{\mu\nu}$ is the flat background metric\footnote{We use the ``mostly plus'' signature for the spacetime metric.}, and $c^a = \psi^\dagger T^a \psi$ is the color charge of the point particle.
Note that \eqref{eq:LagrangianYM} is valid for both $m>0$ and $m=0$.
For massive particles, we set $e(\tau)=1/m$ and $\tau$ will be just the proper time.
In the massless case, we choose an affine parametrization so that $e(\tau)$ is just a constant.
For simplicity we can set $e(\tau)=1$, but keep in mind that $e$ has mass dimension $-1$.
The constraint on the velocity is
\begin{align}
    \bar{g}_{\mu\nu} v^\mu v^\nu = \kappa
    \qquad
    \left\{
    \begin{aligned}
        &\kappa=-1 \quad\mathrm{ for }\quad m>0 \\
        &\kappa=0 \quad\mathrm{ for }\quad m=0.
    \end{aligned}
    \right.
\end{align}
We now consider a static YM field that takes the form of a Kerr-Schild ``single copy'',
\begin{equation}
    \label{eq:KSform}
    A_\mu^a(x) = \frac{g}{4 \pi} \tilde{c}^a \varphi(x) k_\mu(x),
\end{equation}
where $k_\mu(x)$ is a null vector in Minkowski space, $\varphi(x)$ is a scalar field and $\tilde{c}^a$ is the static color charge of the source of the YM field. For a charged test particle moving in this background we crucially require the coupling constant to be small enough to not affect the gauge field configuration. Interestingly, perturbations on a (Coulomb) static YM potential beyond some critical value of the coupling can produce instabilities due to the color charge screening, as was first noticed by Mandula  \cite{Mandula:1976uh,Mandula:1976uh} and then further developed by several authors  \cite{Sikivie:1978sa,Mandula:1977sq,Jackiw:1978zi}.

\begin{figure}[h]
\includegraphics[width=\linewidth]{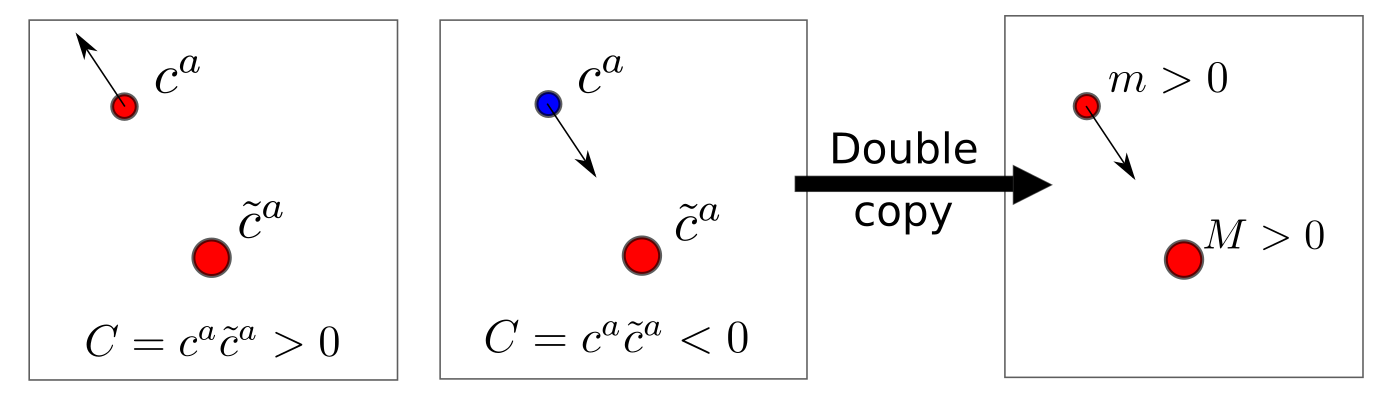}
\caption{For a massive charged particle, we can have both attractive and repulsive gauge theory forces depending on the sign of the charges. This suggests to focus on the case $C < 0$, because masses in gravity are always positive.}
\label{fig:doubleCopy}
\end{figure}

Suppose we have a cyclic coordinate $\xi$, which doesn't appear explicitly in the Lagrangian.
From Noether's theorem,
we know that the corresponding conserved charge is
\footnote{See  \cite{vanHolten:2006xq,Gomis:2021aia} for alternative approaches on how to derive the conserved charges.}
\begin{equation} \label{eq:YMcharge}
    p_\xi^{\mathrm{YM}} \!= \frac{\partial L^{\mathrm{YM}}} {\partial v^\xi} =
    \frac{\partial v^\mu}{\partial v^\xi}
    \!\left( \frac{\bar{g}_{\mu\nu} v^\nu}{e}
    \!-\! \frac{g^2}{4 \pi} c^a \tilde{c}^a \varphi(x) k_\mu(x)  \right)\! ,
\end{equation}
In our convention, the Kerr-Schild double copy relation is,
\begin{align}
    \label{eq:KSdoublecopy}
    \frac{g^2}{4\pi} \to 2 G \qquad
    \tilde{c}^a \to M k_\mu.
\end{align}
The corresponding gravitational field reads,
\begin{equation}
    g_{\mu\nu} = \bar{g}_{\mu\nu} + h_{\mu\nu}
    \quad h_{\mu\nu} = {2 G} M \varphi(x) k_\mu(x) k_\nu(x).
\end{equation}
The flat metric $\bar{g}_{\mu\nu}$ is the same as in the Yang-Mills theory.
The Lagrangian of the point mass in this background reads
\begin{equation}
    \label{eq:LagrangianGR}
    L^{\mathrm{GR}} =  \frac{(\bar{g}_{\mu\nu} + h_{\mu\nu})  v^\mu v^\nu} {2e} - \frac{e m^2}{2},
\end{equation}
where again the einbein is $e=1/m$ for massive particles and $e=1$ for massless particles.
Meanwhile the relativistic constraint is
\begin{align} \label{eq:GRnormal}
    (\bar{g}_{\mu\nu} \!+\!  h_{\mu\nu} ) v^\mu v^\nu = \kappa
    \qquad
    \left\{
    \begin{aligned}
        &\kappa=-1 \quad\mathrm{ for }\quad m>0 \\
        &\kappa=0 \quad\mathrm{ for }\quad m=0.
    \end{aligned}
    \right.
\end{align}
It's clear that $\xi$ is also a cyclic coordinate for $L^{\mathrm{GR}}$, so we have a conserved charge for the point mass
\begin{align} \label{eq:GRcharge}
    p_\xi^{\mathrm{GR}} \!= \frac{\partial L^{\mathrm{GR}}} {\partial v^{\xi}} =
    \frac{\partial v^\mu}{\partial v^{\xi}}
    \left( \frac{\bar{g}_{\mu\nu} v^\nu}{e}
    + \frac{2 G M}{e} \varphi k_\nu v^\nu k_\mu  \right).
\end{align}
From eq. \eqref{eq:YMcharge} and \eqref{eq:GRcharge} we can derive the double copy relation between the conserved charges in Yang-Mills and gravity background by supplementing \eqref{eq:KSdoublecopy} with
\begin{equation}
    \boxed{c^a \to - \frac{v^\mu}{e}
        \quad\text{so that}\quad
        C:= c \cdot \tilde{c} \to - \frac{M}{e} k \cdot v.}
    \label{eq:C_doublecopy}
\end{equation}
We note that the double copy map works for both $C > 0$ and $C < 0$, corresponding to repulsive and attractive forces respectively.
Nevertheless, in the analysis of solutions of the equations of motion, we will focus on the case $C < 0$ to resemble gravity, where the interaction is always ``attractive''(see Fig. \ref{fig:doubleCopy}).

In the case where the dynamics is integrable, knowing the conserved charges is sufficient to fully solve the equations of motion. In particular, this is true for $\sqrt{\text{Schw}}$ and equatorial orbits in $\sqrt{\text{Kerr}}$. In the following sections, we will apply \eqref{eq:C_doublecopy} to obtain the conserved energy and angular momentum for a probe particle moving in the Schwarzschild background and on the equatorial plane of the Kerr background.

\section{Charged test particle in a non-Abelian Coulomb potential}
As proposed in the original work on the black hole double copy  \cite{Monteiro:2014cda}, the single copy of the Schwarzschild solution corresponds to a Coulomb-like potential. In the standard spherical coordinates $x^{\mu}(\tau) = (t, r, \theta, \phi)$, it reads
\begin{align} \label{eq:CoulombField}
A_{t}^a = A_{r}^a =\frac{g}{4 \pi} \frac{\tilde{c}^a}{r}
\quad \rightarrow \quad
F_{r t}^a = - \frac{g}{4 \pi}  \frac{\tilde{c}^a}{r^2},
\end{align}
All the other components of the field strength are vanishing.
\subsection{Massive probe}
The Euler–Lagrange equations of \eqref{eq:LagrangianYM} give us Wong's equations 
\begin{gather}
    \label{eq:Wongeq}
    \frac{d^{2} x^{\mu}}{d \tau^{2}} + \Gamma^\mu{}_{\nu\rho} v^{\nu} v^{\rho}
    = - \frac{g}{m} c^{a} F^{a,\mu}{}_{\nu} v^{\nu} \\
    \frac{d c^{a}}{d \tau} = g f^{a b c} v^{\mu} A_{\mu}^{b} c^{c}(\tau),
\end{gather}
where $\Gamma^\mu{}_{\nu\rho}$ is the Christoffel symbol for spherical coordinates.
Moreover, we have already fixed $v^2=-1$.
Thanks to the spherical symmetry of the problem, we can restrict our analysis to the $x-y$ plane by setting $\theta = {\pi}/{2}$ and $d\theta / d\tau =0$. Then the $\theta-$component of \eqref{eq:Wongeq} is simply
\begin{align} \label{eq:CoulombAngular}
    \frac{d}{d \tau} \left(r^2 \frac{d \phi}{d \tau}\right) = 0 \quad\rightarrow\quad
    L:= r^2 v^{\phi},
\end{align}
which corresponds to the conservation of the $z$-component of the angular momentum $L$. Explicitly, Wong's equations restricted to the $\theta = {\pi}/{2}$ plane are
\begin{equation}
\begin{split}
    \frac{d v^r(\tau)}{d \tau}  - \frac{ L^2}{r(\tau)^3} &= \frac{g^2}{4 \pi m}  \frac{c^a(\tau) \tilde{c}^a}{r(\tau)^2} v^t(\tau)  \\
    \frac{d v^{t}(\tau)}{d \tau} &= \frac{g^2}{4 \pi m} \frac{c^a(\tau) \tilde{c}^a}{r(\tau)^2} v^r(\tau)  \\
    \frac{d c^a(\tau)}{d \tau} &= \frac{g^2}{4 \pi} f^{a b c} v^t(\tau)  \frac{\tilde{c}^b c^c(\tau) }{r(\tau)},
\end{split}
\end{equation}
where we have made manifest the explicit dependence on the proper time $\tau$. A crucial ingredient in solving the equations of motion is to observe that the scalar product of the two color vectors $C := c^{a}(\tau) \tilde{c}^a$ is always conserved
\begin{align}\label{eq:Cconserved}
\frac{d C}{d \tau} &= \frac{g^2}{4 \pi} f^{a b c} v^t(\tau)  \frac{\tilde{c}^a \tilde{c}^b c^c(\tau) }{r(\tau)}  = 0.
\end{align}
In the following, we will consider color charges of opposite sign so that the force is attractive: therefore $C < 0$.
\\
Another conserved charge is the energy which can be defined from the $t-$component of Wong's equation
\begin{align}\label{eq:CoulombEnergy}
\frac{d v^{t}(\tau)}{d \tau} &= -\frac{d}{d \tau} \left[\frac{g^2}{4 \pi m} \frac{c^a(\tau) \tilde{c}^a}{r(\tau)} \right]\rightarrow h := v^{t} + \frac{\alpha}{m} \frac{C}{r},
\end{align}
where for conciseness we define $\alpha = {g^2}/{4 \pi}$.
The energy \eqref{eq:CoulombEnergy} and angular momentum charge \eqref{eq:CoulombAngular} could be derived also directly from the Lagrangian approach as \eqref{eq:YMcharge}.

\begin{figure}[p]
    \includegraphics[scale=0.32]{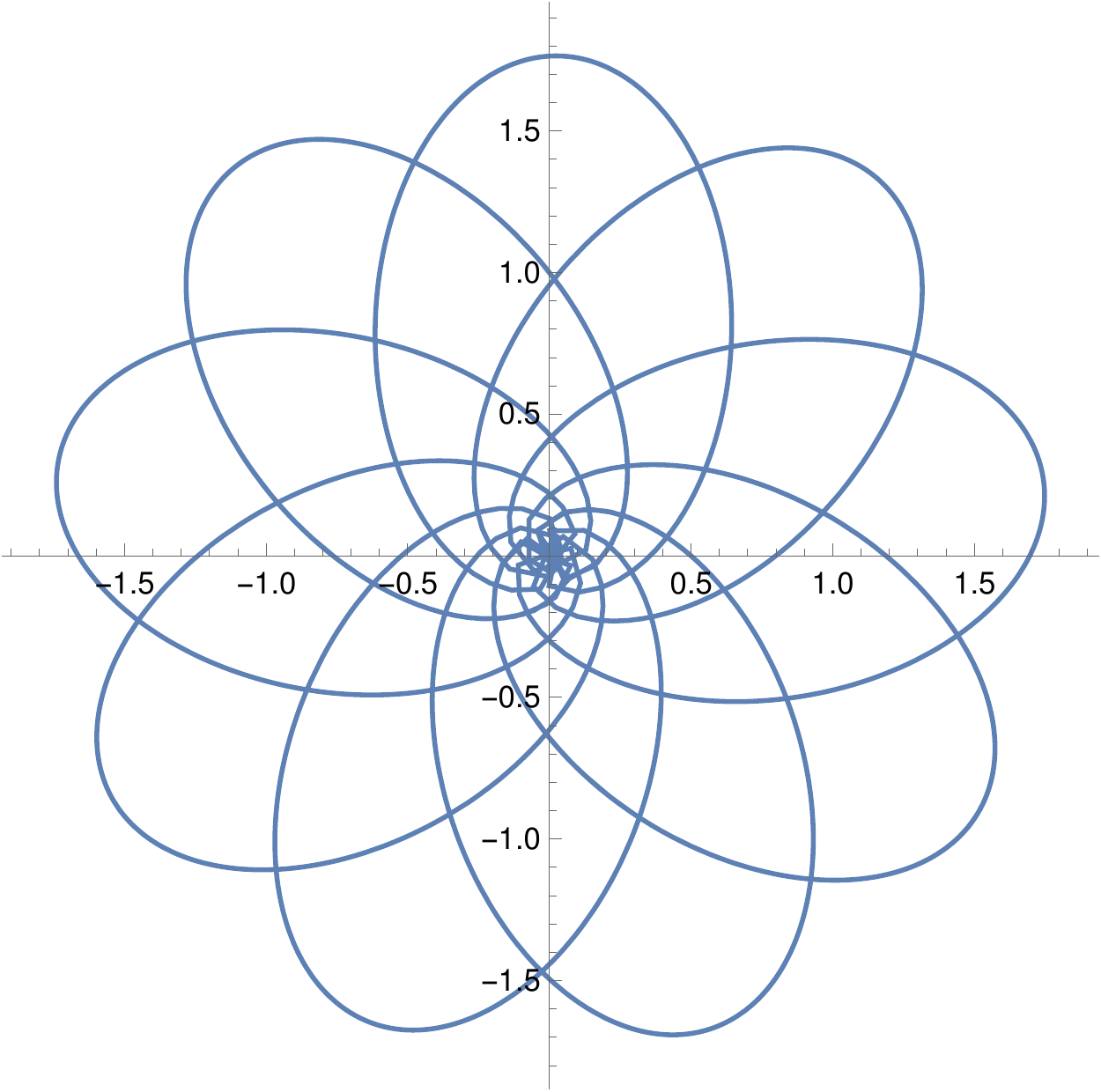}
    \caption{Elliptic orbit for the massive charged particle in \newline \hspace*{2cm} the $\sqrt{\text{Schw}}$ potential. \newline
    }
    \label{fig:ellipticSchw}
\end{figure}

\begin{figure}[p]
    \includegraphics[scale=0.3]{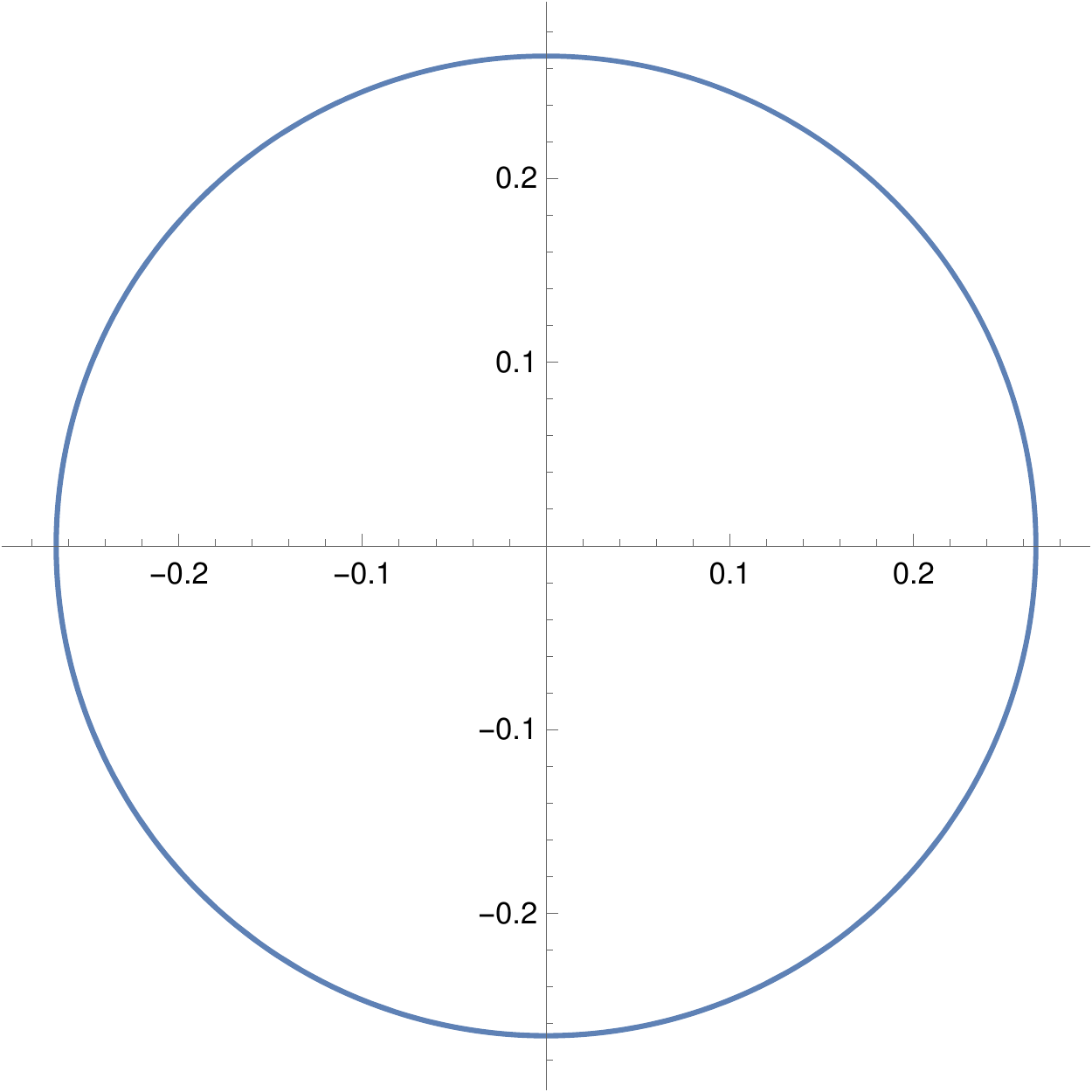}
    \caption{Circular orbit for the massive charged particle in \newline \hspace*{2cm} the $\sqrt{\text{Schw}}$ potential. \newline 
    }
    \label{fig:circularSchw}
\end{figure}

\begin{figure}[p]
    \includegraphics[scale=0.32]{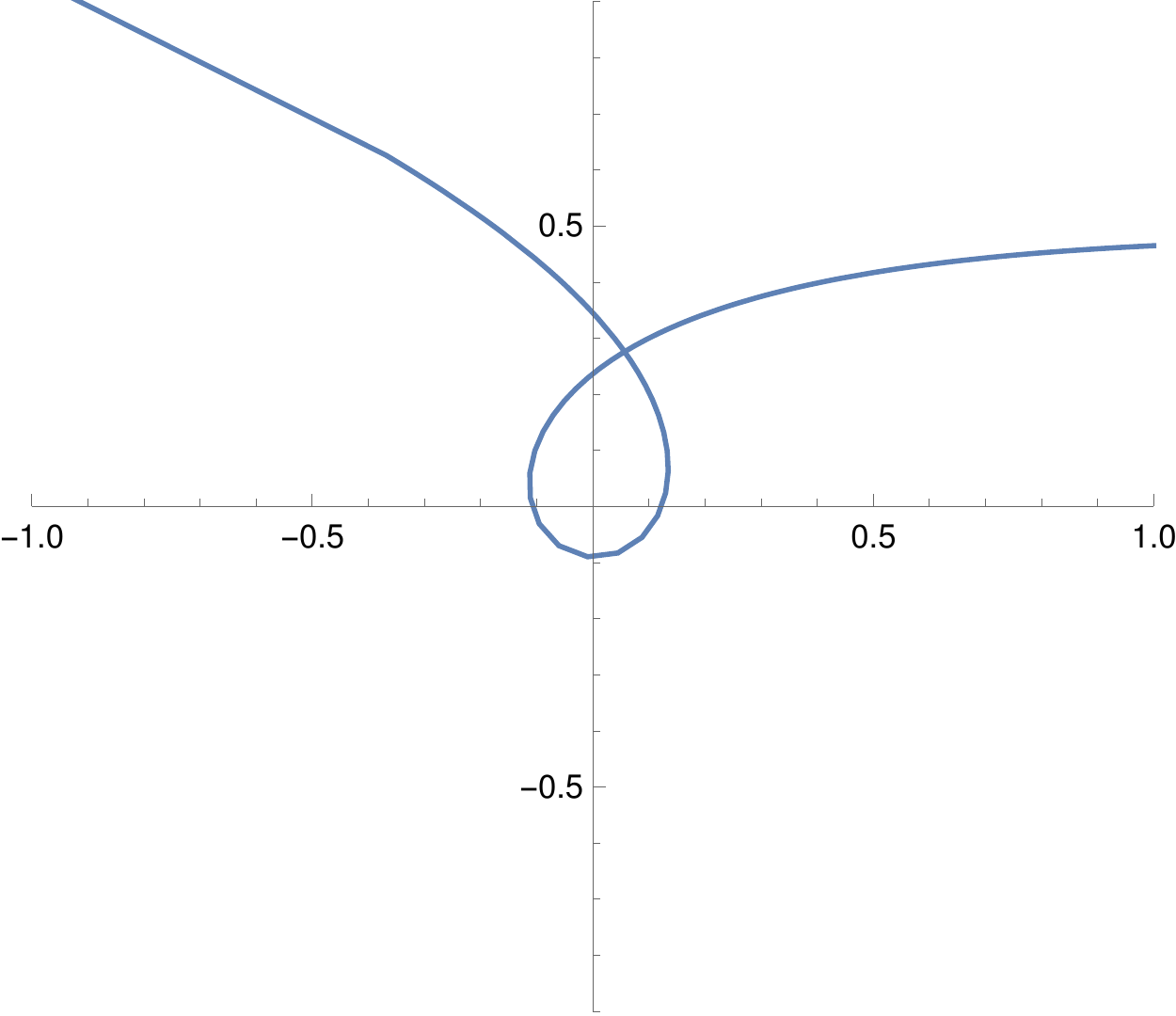}
    \caption{Hyperbolic-type orbit for the massive charged \newline \hspace*{1.5cm} particle in the $\sqrt{\text{Schw}}$ potential. \newline 
    }
    \label{fig:hyperSchw}
\end{figure}

\begin{figure}[p]
    \includegraphics[scale=0.32]{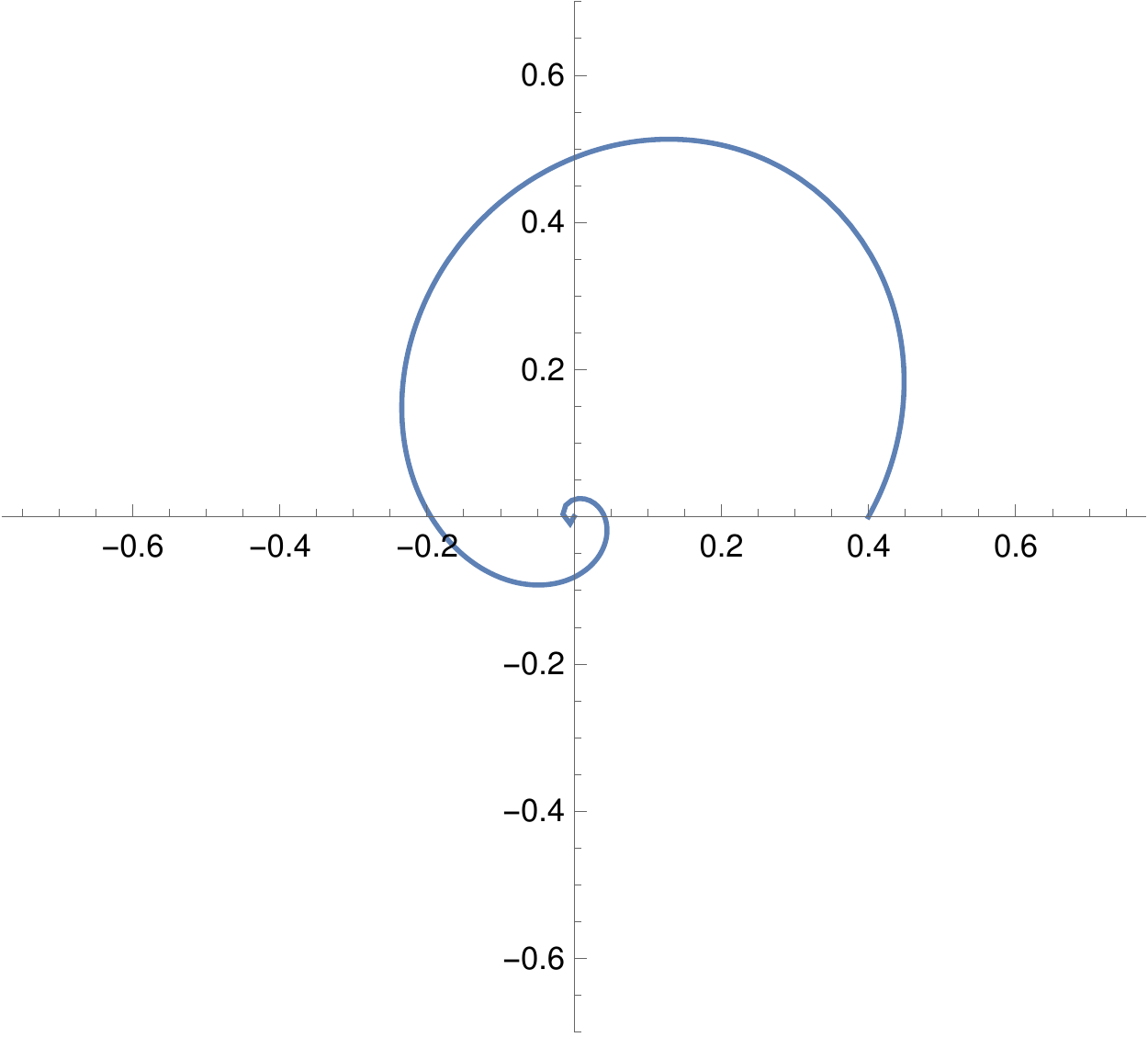}
    \caption{Plunge orbit for the massive charged particle in \newline \hspace*{2cm} the $\sqrt{\text{Schw}}$ potential. \newline 
    }
    \label{fig:plungeSchw}
\end{figure}

Thanks to \eqref{eq:Cconserved}, the non-Abelian Coulomb potential problem can be effectively reduced to the Abelian case, where the strength of the potential is determined by $C$. Even though the structure of the solution is fully known in the Abelian case since long time ago \cite{Landau:1975pou}, we would like to display it here to show the features of the orbits.
Using the $r-$component of the equations of motion we have
\begin{align}
\frac{d^2 r(\tau)}{d \tau^2}  - \frac{ L^2}{r(\tau)^3} &= \frac{\alpha}{m}  \frac{C}{r(\tau)^2} \left(h - \frac{\alpha}{m} \frac{C}{r(\tau)}\right),
\end{align}
and changing variable to $u:= \frac{1}{r}$ as a function of $\phi$
\begin{align}
\frac{d^2 u(\phi)}{d \phi^2} + u(\phi) &= -\frac{\alpha C}{m L^2} \left(h-\frac{\alpha C u(\phi)}{m}\right),
\label{eqn:eom_spinless}
\end{align}
where we have used the simple relation $ \,du / d\phi = -(1/L) dr /d\tau $.
Moreover, $u(\phi)$ is constrained by the relativistic condition $v^{\mu} v_{\mu} = -1$ which effectively reduces the degrees of freedom to the ones of a first order differential equation.
If we define the critical value of the angular momentum
\begin{align}
L_{\text{crit}} = -\frac{\alpha C}{m},
\end{align}
then \eqref{eqn:eom_spinless} can be rewritten as
\begin{align}
\frac{d^2 u(\phi)}{d \phi^2} + u(\phi) &= \frac{L_{\text{crit}}}{L^2} \left(h+ L_{\text{crit}} u(\phi)\right),
\label{eqn:eom_Wong_final_spinless}
\end{align}
while $v^{\mu} v_{\mu} = -1$ gives
\begin{align}
\left(\frac{d u(\phi)}{d \phi}\right)^2 &= \frac{1}{L^2}\left[\left(h+L_{\text{crit}} u(\phi)\right)^2-L^2 u(\phi)^2-1\right].
\label{eqn:eom_Wong_constrain_spinless}
\end{align}
The analytic solution of the differential equation \eqref{eqn:eom_Wong_final_spinless} for $L < L_{\text{crit}}$ is
\begin{align}\label{eqn:plunge_spinless}
    &u^{(\pm)}(\phi) = \frac{B_1 L \sinh \left(\frac{\phi  \sqrt{L_{\text{crit}}^2 \!-L^2}}{L}\right)}{\sqrt{L_{\text{crit}}^2-L^2}}  \\
    &\!+\!\frac{\left(B_2^{(\pm)} \left(L^2 \!-\! L_{\text{crit}}^2\right) \!-\! h L_{\text{crit}}\right) \cosh \Big(\frac{\phi  \sqrt{L_{\text{crit}}^2\!-\! L^2}}{L}\Big) \!+\! h L_{\text{crit}}}{L^2-L_{\text{crit}}^2}, \nonumber
\end{align}
where
\begin{align}
B_1 &= u(\phi)\Bigg|_{\phi=0} \qquad  B_2 = \frac{d u(\phi)}{d \phi} \Bigg|_{\phi=0} \nonumber \\
B_2^{(\pm)} &= \pm \frac{1}{L} \sqrt{\left(h + B_1 L_{\text{crit}}\right)^2-B_1^2 L^2 -1}
\label{eqn:constraint_spinless_massive}
\end{align}
The last equation is directly deduced from eq. \eqref{eqn:plunge_spinless}.
A similar result holds for $L > L_{\text{crit}}$ by using analytic continuation arguments
\begin{align}
    \label{eqn:hyperbolicelliptic_spinless}
    &u^{(\pm)}(\phi) = \frac{B_1 L \sin \left(\frac{\phi  \sqrt{L^2 - L_{\text{crit}}^2}}{L}\right)}{\sqrt{L^2 - L_{\text{crit}}^2}}  \\
    &+\frac{\left(B_2^{(\pm)} \left(L^2\!-\! L_{\text{crit}}^2\right)-h L_{\text{crit}}\right) \cos \Big(\frac{\phi  \sqrt{L^2 \!-\! L_{\text{crit}}^2}}{L}\Big) \!+\! h L_{\text{crit}}}{L^2-L_{\text{crit}}^2}. \nonumber
\end{align}
while the critical case $L = L_{\text{crit}}$ instead gives
\begin{align}
u^{(\pm)}(\phi) = B_2^{(\pm)} + B_1 \phi +\frac{h \phi ^2}{2 L_{\text{crit}}},
\label{eqn:critical_spinless}
\end{align}
where $B_1$ and $B_2^{(\pm)}$ are again given by \eqref{eqn:constraint_spinless_massive}. The presence of a critical value of the angular momentum, which is crucial for the classification of the orbits, is related to the relativistic nature of the problem (see the analysis in \cite{Boyer_2004}). It is convenient to analyze the nature of the orbits by looking at the zeros of the potential, which will be instructive in preparation to the next section. We have
\begin{align}
\left(\frac{d u(\phi)}{d \phi}\right)^2 &= \frac{1}{L^2}\left[\left(h+L_{\text{crit}} u(\phi)\right)^2-L^2 u(\phi)^2-1\right]\nonumber \\
&= \frac{L_{\text{crit}}^2 - L^2}{L^2} (u(\phi) - u_+)(u(\phi) - u_-)\nonumber \\
u_{\pm} &:= \frac{h L_{\text{crit}} \pm \sqrt{\left(h^2-1\right) L^2 + L_{\text{crit}}^2}}{L^2 - L_{\text{crit}}^2}.
\label{eqn:potential_spinless}
\end{align}
where $\frac{d u}{d \phi} \Big|_{u = u_+} = \frac{d u}{d \phi} \Big|_{u = u_-} = 0$.
From this we can deduce that the possible orbits are
\begin{itemize}
\item Elliptic bound orbits (see Fig. \ref{fig:ellipticSchw}) which require two positive roots $u_{\pm} > 0$ with $\frac{d^2 u}{d \phi^2} \Big|_{u = u_{+}} < 0$ and $\frac{d^2 u}{d \phi^2} \Big|_{u = u_{-}} > 0$, that is
\begin{align}
L_{\text{crit}} < |L| &< \frac{1}{\sqrt{1-h^2}}  L_{\text{crit}} \qquad 0 < |h| < 1.
\end{align}
\item Circular orbits (see Fig. \ref{fig:circularSchw}), which correspond to $u_{+} = u_{-} = u_*$ with $\frac{d^2 u}{d \phi^2} \Big|_{u = u_*} = 0$, i.e.
\begin{align}
L &= \pm \frac{1}{\sqrt{1-h^2}}  L_{\text{crit}} \qquad 0 < |h| < 1,
\end{align}
where the radius of such orbits is
\begin{align}
r_*= \frac{1-h^2}{h L_{\text{crit}}}.
\end{align}
\item Hyperbolic-type unbound orbits where the probe escapes to infinity (see Fig. \ref{fig:hyperSchw}), which require just one root to be real and positive $u_- > 0$ with $\frac{d^2 u}{d \phi^2} \Big|_{u = u_-} < 0$
\begin{align}
|L| & >  L_{\text{crit}} \qquad h \geq 1 \nonumber \\
|L| & \leq  L_{\text{crit}} \qquad h \geq 1 \qquad B_2 = B_2^{(+)}
\end{align}
\item Plunge-type orbits for the probe particle (see Fig. \ref{fig:plungeSchw}), provided $u_{\pm} \leq 0$ for $h > 1$ or $u_- > 0 > u_+$ for $0 < h < 1$
\begin{align}
    |L| & \leq  L_{\text{crit}} \qquad 0 < |h| < 1 \nonumber \\
    |L| & \leq  L_{\text{crit}} \qquad h \geq 1 \qquad B_2 = B_2^{(-)}
\end{align}
\end{itemize}
We observe that for $|h|<1$ we always have bound orbits, i.e., the massive particle cannot escape to time-like infinity. This is not surprising since in the limit $r\to \infty$, from eq. \eqref{eq:CoulombEnergy} we have $h=v^t$, which for causality reasons has to be greater than $1$. For the same reason, hyperbolic orbits are allowed only when $h \geq 1$.
\subsection{Massless probe}

In the massless limit, we effectively need to make the replacement
\begin{align}
\alpha/m \to \alpha  \qquad \tau \to \tau'
\label{eqn:rescaling_massless}
\end{align}
to get the new radial equation of motion for the massless charged particle \eqref{eqn:eom_Wong_final_spinless}
\begin{align}
\frac{d^2 u(\phi)}{d \phi^2} + u(\phi) &= \frac{L_{\text{crit}}'}{(L')^2} \left(h'+ L_{\text{crit}}' u(\phi)\right),
\end{align}
with
\begin{align}
h' := v^{t} + \alpha \frac{C}{r} & \qquad L':= r^2 v^{\phi} \nonumber \\
L_{\text{crit}}' :=& -\alpha C
\end{align}
The constraint equation will be,
\begin{align}
\left(\frac{d u(\phi)}{d \phi}\right)^2 &= \frac{1}{(L')^2}\left[\left(h'+L_{\text{crit}}' u(\phi)\right)^2-(L')^2 u(\phi)^2\right].
\label{eqn:constraint_spinless_massless}
\end{align}
and the explicit solution will be given by
\eqref{eqn:plunge_spinless}, \eqref{eqn:hyperbolicelliptic_spinless} and \eqref{eqn:potential_spinless} but with the new constraint equation \eqref{eqn:constraint_spinless_massless} in place of \eqref{eqn:constraint_spinless_massive}.

At this point we can easily study the nature of the solutions. We then have
\begin{itemize}
\item Circular orbits for
\begin{align}
L' &= L_{\text{crit}}' \qquad h' = 0,
\end{align}
with the surprising feature that the radius of the orbit is not constrained.
\item Hyperbolic-type unbound orbits for
\begin{align}
|L'| & >  L_{\text{crit}}' \qquad \text{or} \nonumber \\
|L'| & \leq  L_{\text{crit}}' \qquad B_2 = B_2^{(+)}
\end{align}
\item Plunge behaviour for
\begin{align}
|L'| & \leq  L_{\text{crit}}' \qquad B_2 = B_2^{(-)}
\end{align}
\end{itemize}
These type of orbits are well-represented by the pictures in Fig. \ref{fig:ellipticSchw}, \ref{fig:hyperSchw} and \ref{fig:plungeSchw} respectively. We would like to notice that here, compared to the massive case, elliptic orbits are not allowed and circular orbits are allowed only in the very degenerate limit $h' = 0$.

\subsection{Double copy to Schwarzschild geodesics}

With the equations \eqref{eq:CoulombAngular} and \eqref{eq:CoulombEnergy}, we are now ready to derive the corresponding conserved quantities of a (massive or massless) probe moving on the equatorial plane in a Schwarzschild black hole background.
From eq. \eqref{eq:CoulombField}, we identify the $k_\mu(x)$ and $\varphi(x)$ of Kerr-Schild form \eqref{eq:KSform} as
\begin{align}
    \varphi(x) = \frac{1}{r}, \quad  \qquad k_\mu = (1, 1, 0, 0).
\end{align}
As one can easily check, the Kerr-Schild double copy will give us the Schwarzschild metric in Eddington-Finkelstein coordinates. Applying \eqref{eq:KSdoublecopy} and \eqref{eq:C_doublecopy} to\footnote{We would like to remind the reader that $e$ is the vielbein here.}
\begin{align}
    \label{eq:sqrtSchwhL}
    L^{\sqrt{\mathrm{Schw}}} &= r^2 v^{\phi}  \\
    h^{\sqrt{\mathrm{Schw}}} &= v^{t} + \alpha e \frac{C}{r},
\end{align}
we get
\begin{align}
    \label{eq:SchwL}
    L^{\mathrm{Schw}} &= r^2 v^{\phi} \\
    \label{eq:Schwh}
    h^{\mathrm{Schw}} &= \left(1 - \frac{2 G M}{r} \right) v^t - \frac{2 G M}{r} v^r.
\end{align}
The constraint \eqref{eq:GRnormal} can be written in terms of the conserved charges as
\begin{align}
    \label{eq:Schwv2}
    \left( v^r \right)^2 = \kappa + (h^{\mathrm{Schw}})^2 - \left( 1 + \frac{(L^{\mathrm{Schw}})^2}{r^2} \right)\!
    \left(1- \frac{2GM}{r} \right).
\end{align}
Since the dynamics is integrable, with \eqref{eq:SchwL}, \eqref{eq:Schwh} and \eqref{eq:Schwv2}, one can fully solve the geodesic problem in Schwarzschild. In particular, this implies that the impulse\cite{Arkani-Hamed:2019ymq} and other observables in the probe limit are completely determined by the double copy map.

\section{Charged test particle in a $\sqrt{\text{Kerr}}$ potential}

As proposed in \cite{Monteiro:2014cda}, the single copy of the Kerr solution is given by the following potential
\begin{align}
A_{\mu}^a = \frac{g}{4 \pi} \frac{r^3 \tilde{c}^a}{r^4 + a^2 z^2} \left(1, \frac{r x + a y}{r^2 + a^2},\frac{r y - a x}{r^2 + a^2},\frac{z}{r} \right),
\end{align}
where $r$ is defined implicitly through the following constraint
\begin{align}
\frac{x^{2}+y^{2}}{r^{2}+a^{2}}&+\frac{z^{2}}{r^{2}}=1 \quad \forall (x,y,z) \in \mathbb{R}^3 \!\setminus\! \{x^2 \!+\! y^2 \leq a^2,z=0\} \nonumber \\
r &= 0 \quad \forall (x,y,z) \in \{x^2 + y^2 \leq a^2,z=0\}.
\end{align}
The field is singular on a ring of radius $a$ in the $x-y$ plane, where in fact the source of the field is located. In this section we will always assume $a > 0$.
We will adopt spheroidal coordinates
\begin{align}
x &= \sqrt{r^2 + a^2} \sin(\theta) \cos(\phi)  \nonumber \\
y &= \sqrt{r^2 + a^2} \sin(\theta) \sin(\phi) \nonumber \\
z &= r \cos (\theta ),
\end{align}
which turn the flat metric $\bar{\eta} = \text{diag}(-1,1,1,1)$  into \\$\text{diag}\left(-1, \frac{r^2 + a^2 \cos^2(\theta)}{r^2 + a^2} ,r^2 + a^2 \cos^2(\theta) , (a^2 + r^2)\sin^2(\theta)\right)$.
In the limit where $a\to 0$, this recovers the standard spherical coordinates.
The components of the gauge field in these coordinates are
\begin{align} \label{eq:RootKerrField}
A_t^a &= \frac{g}{4 \pi} \frac{r \tilde{c}^a}{r^2 + a^2 \cos^2(\theta)} \qquad  A_r^a = \frac{g}{4 \pi} \frac{r \tilde{c}^a}{r^2 + a^2}  \nonumber \\
A_{\phi}^a &=  -\frac{g}{4 \pi} \frac{r \tilde{c}^a}{r^2 + a^2 \cos^2(\theta)} a \sin^2(\theta) \qquad A_{\theta}^a = 0.
\end{align}
For simplicity, from now on we will focus on equatorial orbits by setting $\theta = {\pi}/{2}$. We would like to stress that the problem can be solved in full generality, but the complexity is higher for non-equatorial orbits exactly like for Kerr geodesics. The non-vanishing components of the field strength are
\begin{align}
F_{r t}^a = - \frac{g}{4 \pi}  \frac{\tilde{c}^a}{r^2} \qquad F_{r \phi}^a = a \frac{g}{4 \pi}  \frac{\tilde{c}^a}{r^2}.
\end{align}
In the following, we will consider only orbits which lie outside the ring singularity at $x^2 + y^2 = a^2$ on the equatorial plane where \eqref{eq:RootKerrField} is always well-defined.

\subsection{Massive probe}
Wong's equations on the equatorial plane are
\begin{align}
\frac{d v^t}{d \tau} &= \frac{\alpha}{m} \frac{C}{r^2} v^r, \nonumber \\
\frac{d v^r}{d \tau} - \frac{1}{r}(a^2 + r^2) \left(v^{\phi}\right)^2 + &\frac{1}{r} \frac{a^2}{a^2 + r^2}  \left(v^{r}\right)^2 \nonumber \\
&\hspace{-10pt}= \frac{\alpha}{m} \frac{C}{r^4} \left[\left(v^t - a v^{\phi}\right) \left(a^2 + r^2\right) \right], \nonumber \\
\frac{d v^{\theta}}{d \tau} &= 0, \nonumber \\
\frac{d v^{\phi}}{d \tau} + 2 \frac{r}{r^2 + a^2} v^{r} v^{\phi}&= a \frac{\alpha}{m} \frac{C}{r^2 (r^2 + a^2)} v^r.
\label{eqn:eom_spin}
\end{align}
As in the Coulomb potential case, there is a notion of conserved energy and angular momentum
\begin{align} \label{eq:chargeRootKerr}
    h &:= v^{t} + \frac{\alpha}{m} \frac{C}{r}  \\
    L &:= (r^2 + a^2) v^{\phi} + a \frac{\alpha}{m} \frac{C}{r}.
\end{align}
In particular, the angular momentum now also includes contributions from the parameter $a$ which is perfectly analogous to the Kerr case.

At this point we can use those conserved charges to derive the equation of motion for $u(\tau)$ using \eqref{eqn:eom_spin}. We find
\begin{align}
\frac{d v^r}{d \tau}&-\frac{(L r + a (L_{\text{crit}} - r v^r))(L r + a (L_{\text{crit}} + r v^r))}{r^3 (a^2 + r^2)} \nonumber \\
&\hspace{25pt}= - \frac{L_{\text{crit}}}{r^4} \left[a(a h-L)+r (L_{\text{crit}} + h r)\right].
\label{eqn:eom_spin_general}
\end{align}
The constrained equation for $v^r$ gives
\begin{align}
\left(v^r\right)^2 =& \left(1 + \frac{a^2}{r^2} \right) (h - V_+) (h - V_-) - \left(1 + \frac{a^2}{r^2} \right) \nonumber \\
\hspace{-10pt}V_{\pm} :=& \frac{1}{r} \left[-L_{\text{crit}} \pm \frac{|L r + a L_{\text{crit}}|}{\sqrt{r^2 + a^2}} \right]
\label{eqn:Veff_spin}
\end{align}
It is worth noticing here that the leading term in $1/r$ on the RHS of \eqref{eqn:eom_spin_general} is proportional to $a$ and to $a h - L$, which is a feature shared also by Kerr equatorial orbits \cite{Chandrasekhar:1985kt}. \\
In the limit $a \to 0$, the solution collapses to the Coulomb non-abelian potential we already considered in the last section because
\begin{align}
V_{\pm} \stackrel{a = 0}{\to}& \frac{1}{r} (-L_{\text{crit}} \pm |L| )
\end{align}
Regarding the case $L = a h$, an analytic solution for $\tau$ as a function of $r$ exists (similarly to Kerr black hole equatorial geodesic) but we will not display it here.

A pressing question is whether circular orbits exist, and what are the corresponding values for the energy and the angular momentum in these cases. The condition for the existence of circular orbits at $r=r_*$ (meaning that the radius is $\sqrt{r_*^2+a^2}$) is given by the common solution of
\begin{align}
v^r|_{r=r_*} = 0 \qquad \left. \frac{d v^r}{d r} \right|_{r=r_*}\!\!= 0.
\end{align}
With some algebra we can reduce this system of equations to
\begin{align}
h^2&=1 - \frac{L_{\text{crit}}}{r_*^3} \left(a x + h r_*^2\right) \qquad x:= L - a h \nonumber \\
\left(a^2-x^2\right)^2 &= -\frac{L_{\text{crit}}}{r_*^3} \left(a^2 +r_*^2\right) \left(x^2+ r_*^2\right) (4 a x -L_{\text{crit}} r_*),
\label{eqn:x_eq}
\end{align}
where the last equation is a quartic polynomial in $x$. As proved in Appendix \ref{se:AppendixB}, for every $a > 0$ there are two distinct real (and therefore other two complex) solutions for \eqref{eqn:x_eq}. We will call the real roots $x_1$ and $x_2$, and we order them as $x_1 > x_2$. The value of the energy and the angular momentum for such circular orbits is given by \eqref{eqn:x_eq}, i.e. explicitly
\begin{align}
h^{\pm}_{1,2}&= \frac{1}{2 r_*} \left(-L_{\text{crit}} \pm \sqrt{L_{\text{crit}}^2 +4 r_*^2 -4 \frac{a}{r_*} L_{\text{crit}} x_{1,2}} \right) \nonumber \\
L^{\pm}_{1,2}&= x_{1,2} + a h^{\pm}_{1,2}.
\label{eq:massiveCircular}
\end{align}
where only $(h^{+}_{1},L^{+}_{1})$ always satisfies the causality constraint. As we we will see later, this solution will indeed be related with stable circular orbits.

\begin{figure}[h!]
    \includegraphics[scale=0.3]{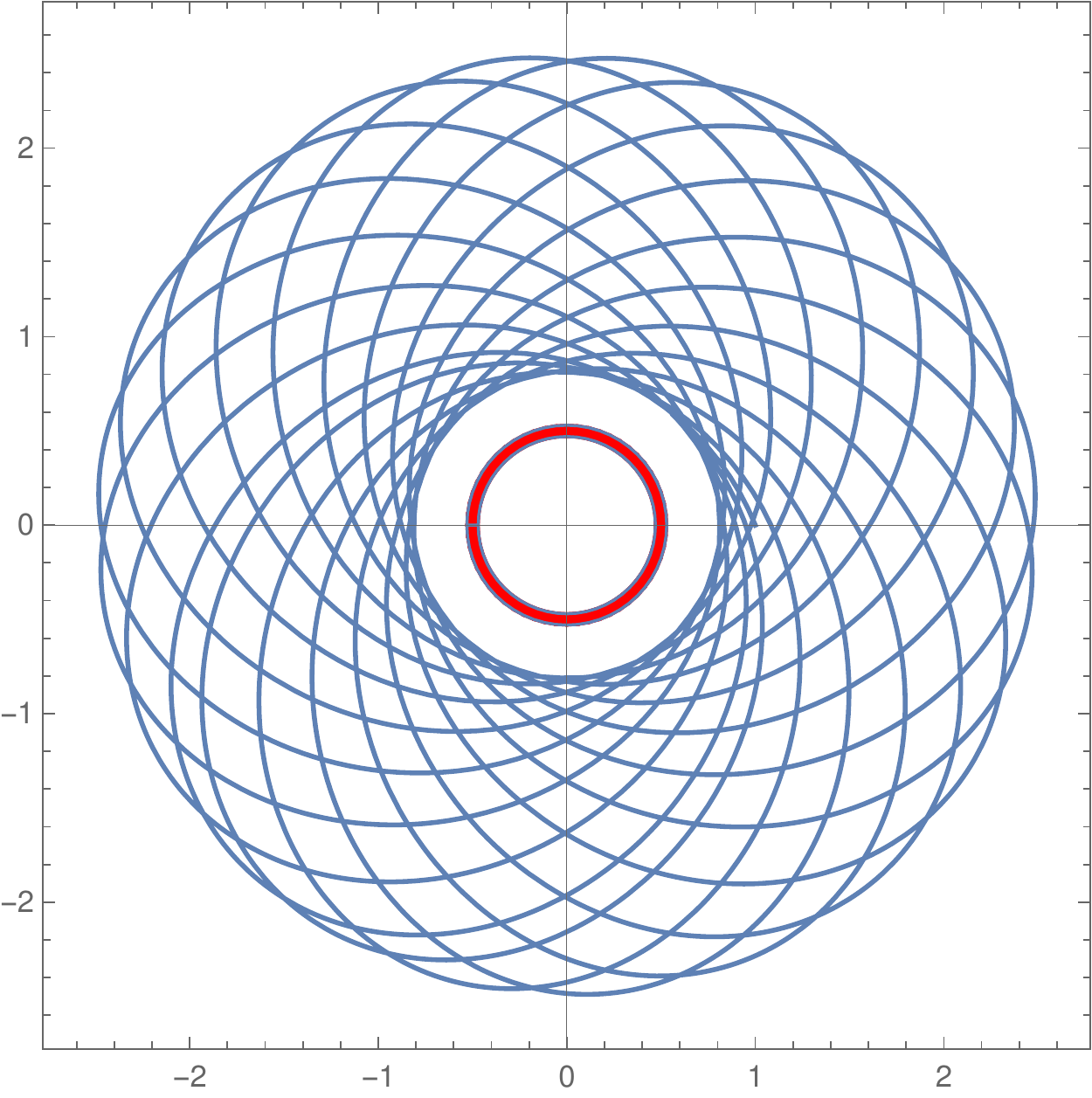}
    \caption{Elliptic orbit for the massive charged particle in \newline \hspace*{2cm} the $\sqrt{\text{Kerr}}$ potential. \newline 
    }
    \label{fig:ellipticKerr}
\end{figure}

\begin{figure}[h!]
    \includegraphics[scale=0.3]{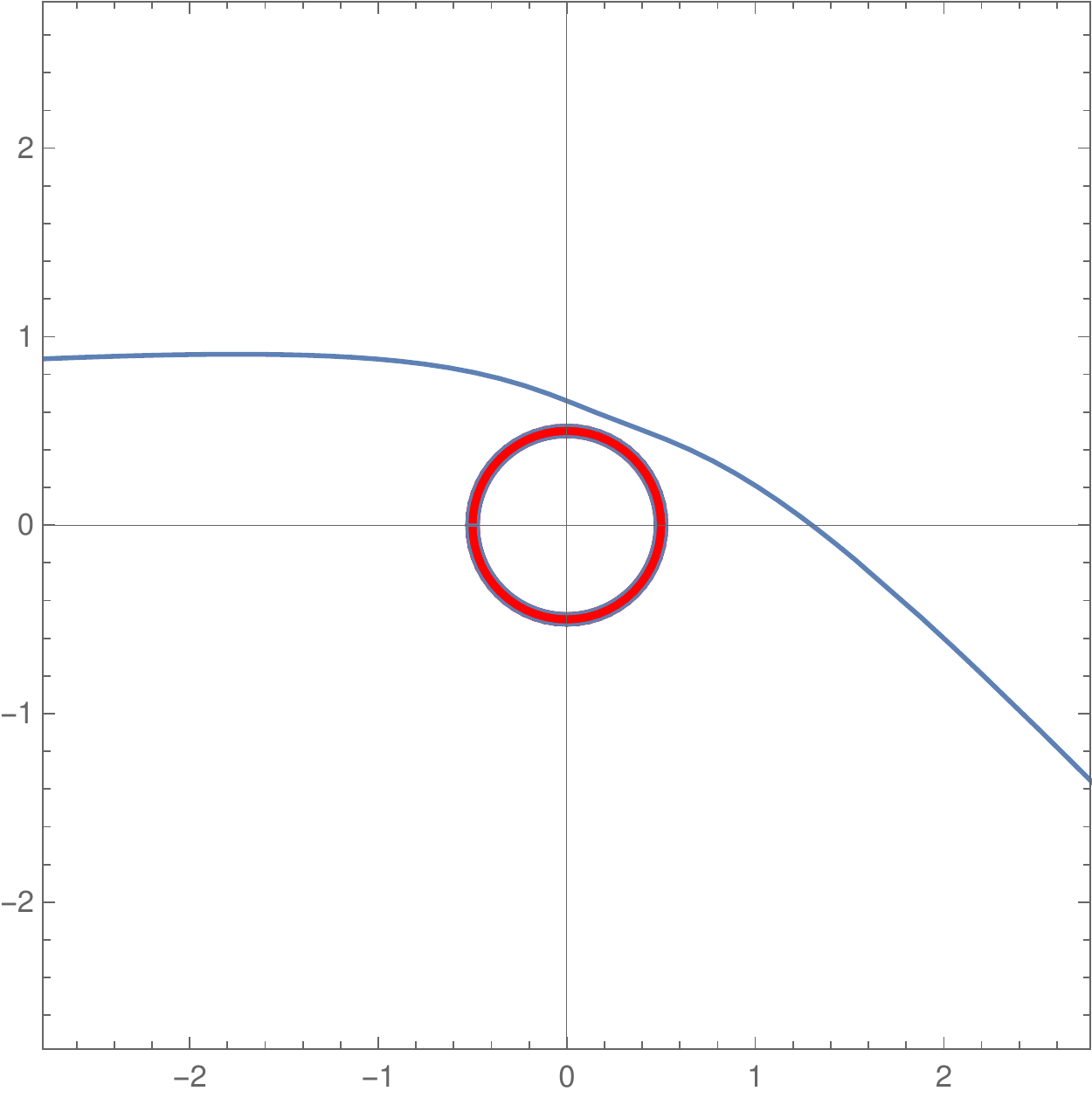}
    \caption{Hyperbolic-type orbit for the massive charged \newline \hspace*{1.5cm} particle in the $\sqrt{\text{Kerr}}$ potential. \newline 
    }
    \label{fig:hyperKerr}
\end{figure}

\begin{figure}[h!]
    \includegraphics[scale=0.3]{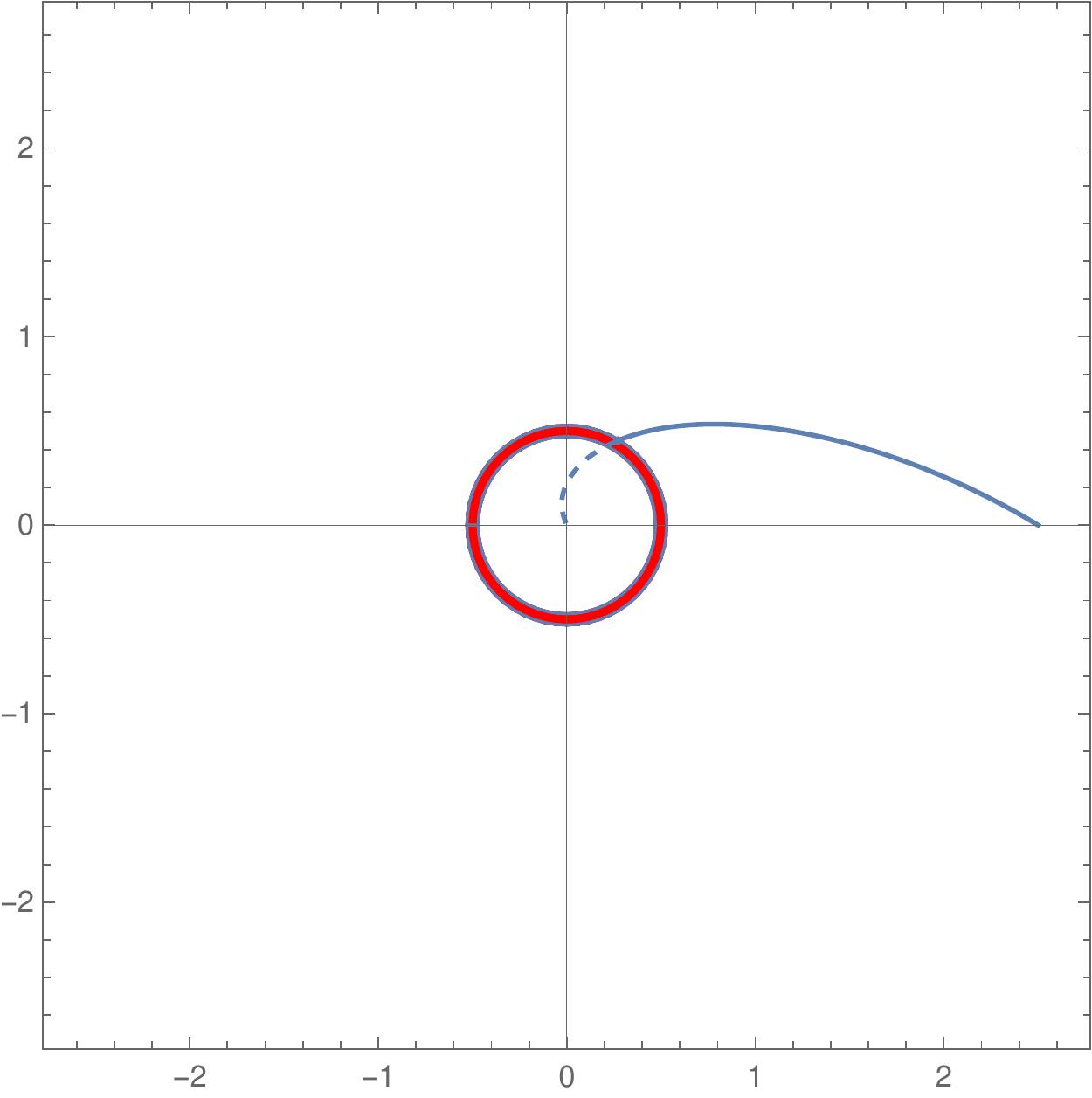}
    \caption{Plunge orbit for the massive charged particle in the \newline \hspace*{2cm} $\sqrt{\text{Kerr}}$ potential. \newline 
    }
    \label{fig:plungeKerr}
\end{figure}

In order to study the general case, we need to analyse the nature of the roots of the RHS of the constraint equation \eqref{eqn:Veff_spin}.
Specifically, we have
\begin{align}
\left(v^r\right)^2 =& \frac{1}{r^3} \mathcal{P}(r) \nonumber \\
\mathcal{P}(r) :=& \left(h^2-1\right) r^3 +2 h L_{\text{crit}} r^2  \\
&+ \left(a^2 \left(h^2-1\right)+L_{\text{crit}}^2-L^2\right) r + 2 a L_{\text{crit}} (a h-L) \nonumber
\end{align}
defines a third order polynomial $\mathcal{P}(r)$, to which we can apply the tools developed in Appendix \ref{se:AppendixB} in order to understand the nature of the roots. By computing the reduce discriminant $\Delta_R (\mathcal{P}(r))$ (see \eqref{eqn:detr}), we can establish whether
\begin{itemize}
\item $\mathcal{P}(r)$ has three simple real roots ($\Delta_R (\mathcal{P}(r)) > 0$)
\item $\mathcal{P}(r)$ has one simple real root ($\Delta_R (\mathcal{P}(r)) < 0$)
\item $\mathcal{P}(r)$ has a double or triple root ($\Delta_R (\mathcal{P}(r)) = 0$)
\end{itemize}
and by using Descartes' rule of signs we can also understand the number of positive/negative roots.
We note that in the case $\Delta_R (\mathcal{P}(r)) = 0$ there are circular orbits with radius given by solving eq. \eqref{eqn:x_eq}.
However, $\mathcal{P}(r)$ by itself is a polynomial in $h$ of degree $8$, which is not possible to solve analytically. Therefore, we can only qualitatively analyse the real solutions of $\mathcal{P}(r)=0$. We find that the properties of the solution depend on the values of $L_{\mathrm{crit}}$ and $L$. Specifically, we find exactly two cases:
\begin{itemize}
    \item Case 1) $L_{\mathrm{crit}}<a$: for a given value of $L$ there are at most $4$ real solutions for $h$, and only two of them can be positive. We denote them as $h_A, h_B$ with $h_A<h_B$ when they exist.
    \item Case 2) $L_{\mathrm{crit}}>a$: for a given value of $L$, there are either $2$ or $4$ real solutions for $h$. One of these real solutions is $h_{1}^{+}$, which is defined by \eqref{eq:massiveCircular}.
\end{itemize}
In addition, as discussed before $h=L/a$ represents another critical value of the energy since having $h > L/a$ or $h < L/a$ will change of the sign of the constant term in the polynomial $\mathcal{P}(r)$.
A detailed analysis for the orbits shows that we can have the following cases
\begin{itemize}
\item Elliptic orbits for\footnote{When $L_{\text{crit}} = a$, $h_B$ becomes equal to $L/a$.}
\begin{align}
h_A < h < \text{min}\{h_B,1\} \qquad &L_{\text{crit}} < a  \nonumber \\
h_{1}^{+} < h < \text{min}\{L/a,1\} \qquad &L_{\text{crit}} > a  \nonumber
\end{align}
\item Hyperbolic-type orbits for $h > 1$ in all cases
\item Plunge behaviour for $h > L/a$ in all cases
\end{itemize}
where $h_A$ can be identified with the stable circular orbit value $h_1^+$ and it is understood that when two intervals overlap we can have different types of orbits according to the boundary conditions (like the sign of the initial velocity or also the initial radial coordinate). We have represented the typical behaviour of those solutions in Fig. \ref{fig:ellipticKerr}, Fig. \ref{fig:hyperKerr} and Fig. \ref{fig:plungeKerr}, where we have also highlighted in red the ring singularity at $r = 0$ where the gauge potential has a singular behaviour.

\begin{figure}[h!]
    \includegraphics[scale=0.3]{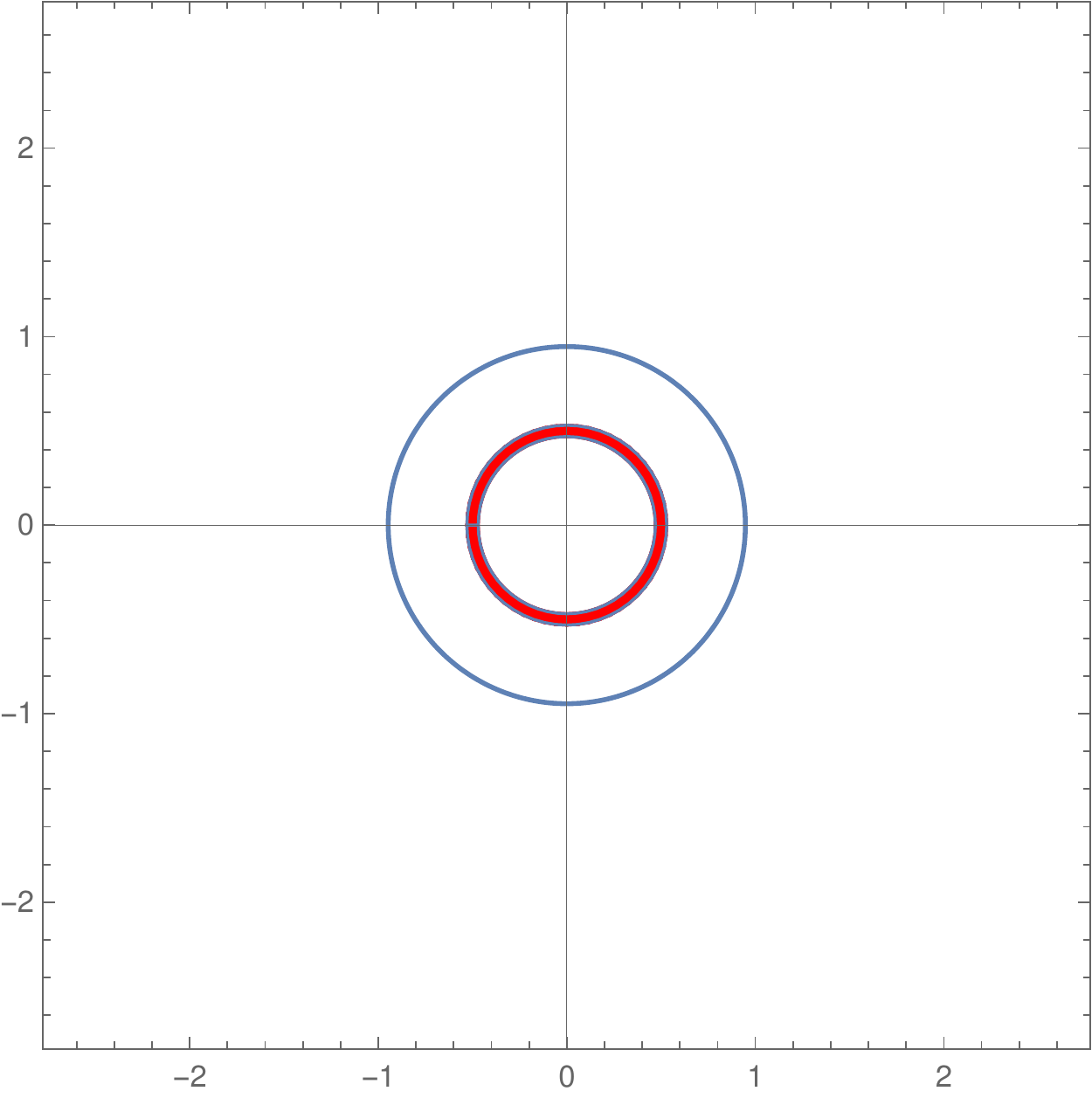}
    \caption{Circular orbit for the massive charged particle in \newline
    \hspace*{2cm} the $\sqrt{\text{Kerr}}$ potential. \newline 
    }
    \label{fig:circularKerr}
\end{figure}

An interesting limit is the one which correspond to marginally bound circular orbits with $h = 1$ (see Fig. \ref{fig:circularKerr}): in such case we can find a simple analytic solution for the value of the radius and of the charges
\begin{align}
L_{\pm} \Big|_{h = 1} &= a-\left(\sqrt{a} \pm \sqrt{L_{\text{crit}}} \right)^2 \nonumber \\
r_{*,\pm} \Big|_{h = 1} &= a \pm \sqrt{a L_{\text{crit}}},
\end{align}
where $r_{*,-}$ exists only when $L_{\text{crit}} < a$.
Since our main goal is to connect the conserved charges on the gauge side with the ones on the gravity side, we leave a full analytic analysis of the generic orbits for massive particles in the case $a > 0$ for a future study.

\subsection{Massless probe}

Using \eqref{eqn:rescaling_massless}, we can derive the corresponding equations of motion for massless charged test particle in the $\sqrt{\text{Kerr}}$ potential. In particular the relativistic constraint equation becomes
\begin{align}
\left(v^r\right)^2 =& \left(1 + \frac{a^2}{r^2} \right) (h' - V'_+) (h' - V'_-)
\label{eqn:Veff_spin_massless}
\end{align}
where the potential and the conserved quantities in the massless case are
\begin{align}
V'_{\pm} :=& \frac{1}{r} \left[-L_{\text{crit}}' \pm \frac{|L' r + a L_{\text{crit}}'|}{\sqrt{r^2 + a^2}} \right] \\
h' := v^{t} +& {\alpha} \frac{C}{r}
 \qquad\quad L_{\text{crit}}' := -\alpha C \\
L' :=& (r^2 + a^2) v^{\phi} + \frac{\alpha a C}{r}.
\end{align}
Causality requires $v^t>0$ and therefore the conserved energy has to satisfy $h'>-L_{\mathrm{crit}}'/r$. Meanwhile, from \eqref{eqn:Veff_spin_massless} we know the region $V'_{-}<h'<V_{+}'$ is forbidden otherwise the right-hand side is negative.
Since $V'_{-}\leq -L_{\mathrm{crit}}'/r \leq V_{+}'$, the physically meaningful value of the energy is constrainted as $h'\geq V_{+}'$.
Taking the time-derivative of \eqref{eqn:Veff_spin_massless} we can derive
\begin{align}
\frac{d v^r}{d \tau} =& - \left(1 + \frac{a^2}{r^2}\right) \left[\frac{d V_+'}{d r} (h' - V_-') + \frac{d V_-'}{d r} (h' - V_+') \right] \nonumber \\
&- \frac{a^2}{r^3} (h' - V'_+) (h' - V'_-)
\end{align}
and therefore the condition to have circular orbits at $r=r_*$ is equivalent to
\begin{align}
h'|_{r=r_*} =& V'_{+}|_{r=r_*} \qquad \left. \frac{d V'_+}{d r} \right|_{r=r_*}\! = 0
\end{align}
We find that a circular orbit requires the energy and angular momentum to satisfy
\begin{align}
    h'_{\pm} &= \frac{a^2 L_{\text{crit}}'}{r_*^3} \left(1 \pm \sqrt{1 + \frac{r_*^2}{a^2}} \right)  \\
    L'_{\pm} &= -\frac{a^3 L_{\text{crit}}'}{r_*^3}
    \left( 1 + \frac{2r_*^2}{a^2} \pm \left(1 + \frac{r_*^2}{a^2}\right)^{3/2}\right)
\end{align}
As it is easy to show, in the vicinity of the singular ring $r \to 0$, we have
\begin{align}
\lim_{r \to 0} V'_+(r) = \frac{L'}{a}
\end{align}
A detailed analysis for the orbits show that we can have
\begin{itemize}
\item Hyperbolic-type orbits for any positive value of the energy $h' > 0$.
\item Plunge behaviour for $h' \geq L'/a$ both in the co-rotating case $L' > 0$ and in the in the counter-rotating case $L' < 0$.
\item Elliptic orbits for $h_{-}' < h' < \text{min}\{0,L'/a\}$ with  $L' < L_{\mathrm{crit}}$. $h'_{-}$ is determined implicitly in terms of the minimum of $V'_+$.
\item Stable circular orbits for $(h',L') = (h'_{-}, L'_{-})$ with $L<L_{\mathrm{crit}}$ and unstable circular orbits for $(h',L') = (h'_{+},L'_{+})$ with $L<-L_{\mathrm{crit}}$.
\end{itemize}
where when two regions of the parameter space $(h',L')$ overlap we can have different types of orbits according to the initial boundary conditions. This behaviour of the massless probe particle is also (at least qualitatively) exemplified by the pictures in Fig. \ref{fig:ellipticKerr}, Fig. \ref{fig:hyperKerr}, Fig. \ref{fig:plungeKerr} and Fig. \ref{fig:circularKerr} of the previous section. Unlike null geodesics on the Kerr background, elliptic orbits are surprisingly allowed in the $\sqrt{\mathrm{Kerr}}$ case and there are stable circular orbits.

\subsection{Double copy to Kerr geodesics on equatorial orbits}

We can now obtain the conserved charges for a probe particle moving in the Kerr black hole background. In the Kerr-Schild form we have
\begin{align}
    &\varphi(x) = \frac{r}{a^2 \cos^2(\theta )+r^2} \\
    &k_\mu = \left( 1, \frac{r^2 + a^2 \cos^2(\theta )}{a^2+r^2}, 0, -a \sin ^2(\theta )\right).
\end{align}
The conserved quantities in the Kerr spacetime is gained from \eqref{eq:KSdoublecopy} and \eqref{eq:C_doublecopy}, i.e. from
\begin{align}
    \label{eq:sqrtKerrhL}
    L^{\sqrt{\mathrm{Kerr}}} &= (r^2 + a^2) v^{\phi} + a \alpha e \frac{C}{r}  \\
    h^{\sqrt{\mathrm{Kerr}}} &= v^{t} + \alpha e \frac{C}{r},
\end{align}
we get
\begin{align}
\label{eq:KerrhL}
    \hspace{-8pt}L^{\mathrm{Kerr}}  = &(r^2 + a^2) v^{\phi} - a \frac{2G M}{r} \left( \frac{r^2 v^r}{a^2+r^2}-a v^{\phi }+v^t \right) \\
    \hspace{-8pt}h^{\mathrm{Kerr}}  = &v^{t} - \frac{2GM}{r} \left( \frac{r^2 v^r}{a^2+r^2}-a v^{\phi }+v^t \right).
\end{align}
The null-like and time-like geodesics on the equatorial orbit can then be obtained by using these two charges and the four-velocity normalization condition \eqref{eq:GRnormal}.

\section{Conclusions}

The color-kinematics duality offers promising ideas to tackle complex problems in the gravitational setting, in particular regarding the two-body problem for two massive particles in general relativity. In the extreme limit where one mass is much bigger than the other, i.e. at leading order in the expansion in the mass ratio, the problem is equivalent to a light particle following geodesics in the background sourced by the other heavy particle. Making use of the Kerr-Schild double-copy, one can derive the background metric on which the probe massive particle move from the corresponding gauge field configuration. In particular, for Schwarzschild and Kerr, those correspond to a non-Abelian ${1}/{r}$ Coulomb-like potential ($\sqrt{\text{Schw}}$) and to the potential generating by a rotating disk of charge ($\sqrt{\text{Kerr}}$) \cite{Monteiro:2014cda}. The latter can be also formally derived from the former using the Newman-Janis shift \cite{Newman:1965tw,Arkani-Hamed:2019ymq,Guevara:2020xjx}.

In this work, we consider a test charged probe particle moving in the $\sqrt{\text{Schw}}$ and the equatorial plane of $\sqrt{\text{Kerr}}$ potential and we solve Wong's equations in terms of the conserved energy $h$ and the angular momentum $L$. 
In particular we focus on the case where the color charge $C = c^a \tilde{c}^a$ is negative, so that we can correctly reproduce similar orbits with gravity, which is always attractive. 
We can then extend the Kerr-Schild double-copy to derive a mapping between conserved charges of a probe particle in the YM and in the gravitational background. Specifically, the map \eqref{eq:C_doublecopy} replaces the color charge of the test particle by its momentum in the spirit of color-kinematics duality. This allows us not only to recover fully the geodesic equations for Schwarzschild and Kerr, but also provides the bridge with the perturbative double copy prescription for charged particles introduced by Goldberger and Ridgway to relate the gluon and the graviton radiative field \cite{Goldberger:2016iau}. We believe that a correspondence of our charge double copy with the standard BCJ double copy can be done. For example, we can compare some set of observables in the classical limit --like the momentum impulse -- computed for an unbound orbit of a test particle with standard scattering amplitude techniques (i.e. expanding around a straight-line trajectory). While the double copy was originally discovered in the S-matrix formalism and therefore for unbound-like orbits, our mapping applies naturally also for bound problems \cite{Goldberger:2017vcg}. This is because explicit solutions of equations of motion do have a natural analytic continuation in terms of the conserved charges \cite{Goldberger:2017vcg, Kalin:2019rwq}.

The $\sqrt{\text{Schw}}$ and $\sqrt{\text{Kerr}}$ potential are of interest on their own, in particular to understand better the YM dynamics in a non-perturbative setting. Indeed, stability of these type of potentials have been investigated by Mandula et al. long time ago in relation to the confinement mechanism \cite{Mandula:1976uh,Mandula:1976uh,Sikivie:1978sa,Mandula:1977sq,Jackiw:1978zi}. For our work, we keep the coupling constant small enough so that the probe particle does not affect the gauge background. While for $\sqrt{\text{Schw}}$ we have found an analytical solution for any type of orbit, for $\sqrt{\text{Kerr}}$ we have discussed qualitatively the behaviour of the probe particle moving in equatorial orbits. For both potentials, we have found that for massive test particles can move in elliptic, circular, hyperbolic-type or plunge orbits depending on the values of the conserved charges. For massless particles the situation is similar, but with a surprise: while elliptic orbits are not allowed and circular orbits become unstable in $\sqrt{\text{Schw}}$, there are instead elliptic and stable circular orbits for $\sqrt{\text{Kerr}}$. With this exception in mind, what is striking to us is the similarity of those solutions for both backgrounds with the behaviour of time-like and null-like geodesics of Schwarzschild/Kerr \cite{Chandrasekhar:1985kt}, which is evident both from intermediate stage calculations and also from the explicit analytic results. It would be very interesting to extend our analysis to generic non-equatorial orbits in Kerr, perhaps by looking for a gauge theory analogue of the Carter constant \cite{Carter:1968rr}.

Summarizing, the probe limit contains many information on full two-body problem in general relativity \cite{Damour:2016gwp,Cheung:2020gbf,Cheung:2020gyp}, and our results provide another indication that such data is entirely encoded in the simpler gauge theory dynamics via the double copy map. This is supported also by previous evidence coming from the derivation of the impulse \cite{Arkani-Hamed:2019ymq} and the multipoles \cite{Chacon:2021hfe} using double copy techniques. While we have considered only the leading order contribution in the so-called self-force expansion, it would be nice to understand whether double copy can help to shed light also on the higher order terms in the expansion in the mass ratio. We leave this interesting problem for a future discussion.

\acknowledgments

We thank Donal O'Connell and Jan Plefka for very helpful discussions and comments on the draft. We would like to thank the organizers of the GGI program ``Gravitational scattering, inspiral, and radiation'' and of the AEI workshop ``Rethinking the Relativistic Two-Body Problem'' for providing a stimulating environment, where part of these ideas were developed. This project has received funding from the European Union's Horizon 2020 research and innovation program under the Marie Sklodowska-Curie grant agreement No.764850 ``SAGEX''.
\newpage
\appendix

\section{Shape of the $\sqrt{\text{Kerr}}$ potential}
\label{se:AppendixA}

We would like to show here the structure of the time-component of the potential $A_{t}$ for $\sqrt{\text{Kerr}}$, projected along the $x-z$ plane (there is always an azimuthal symmetry). While for $a > 0$ at large distances from the singularity $A_{t}$ has an ellipsoidal shape, closer to the singularity line of width $a$ the potential develops a dipole-type configuration.

\begin{figure}[!h]
    \centering
    \begin{minipage}{0.25\textwidth}
        \centering
        \includegraphics[scale=0.30]{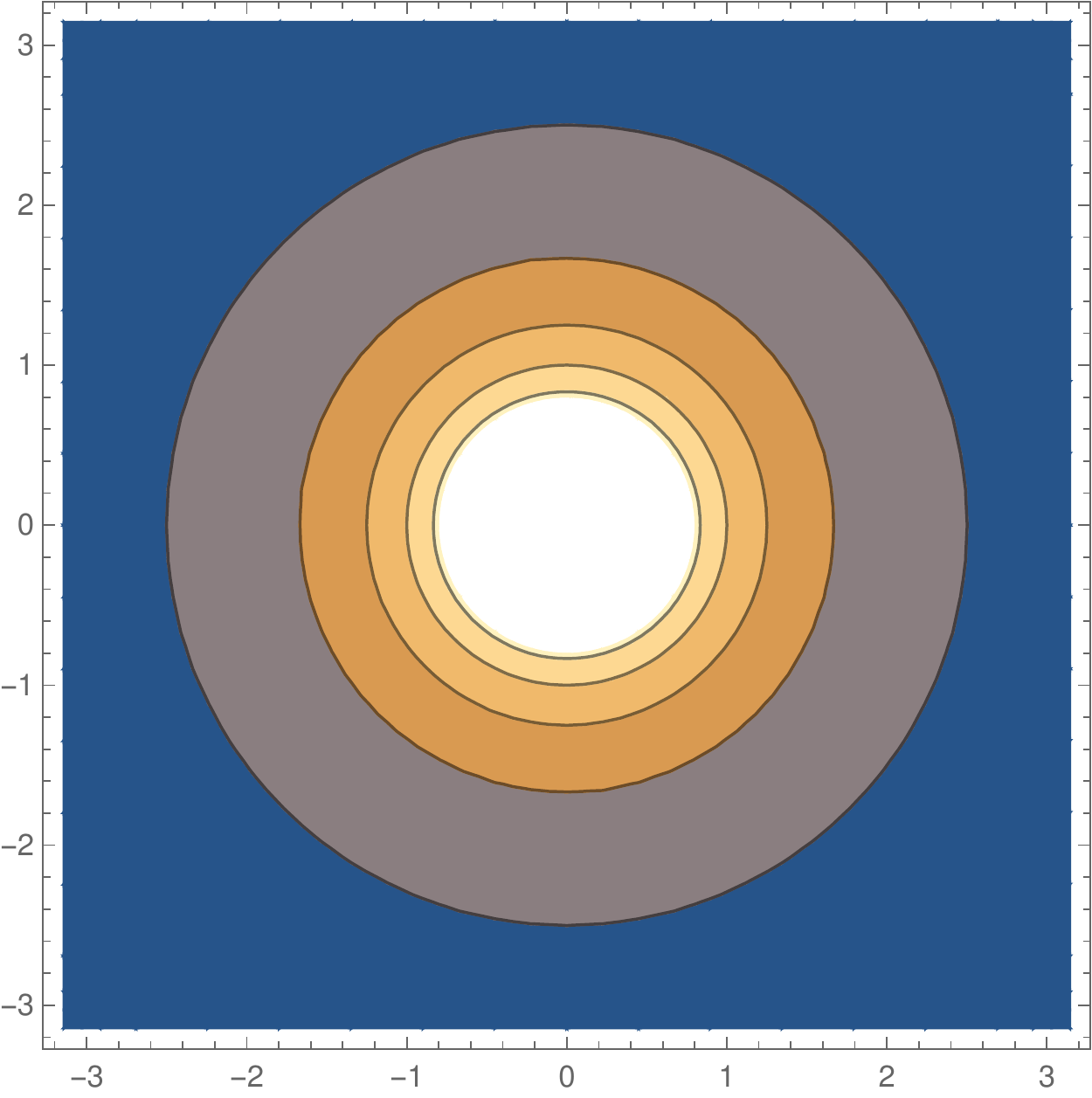}
        \label{fig:potential1}
    \end{minipage}%
    \begin{minipage}{0.25\textwidth}
        \centering
        \includegraphics[scale=0.30]{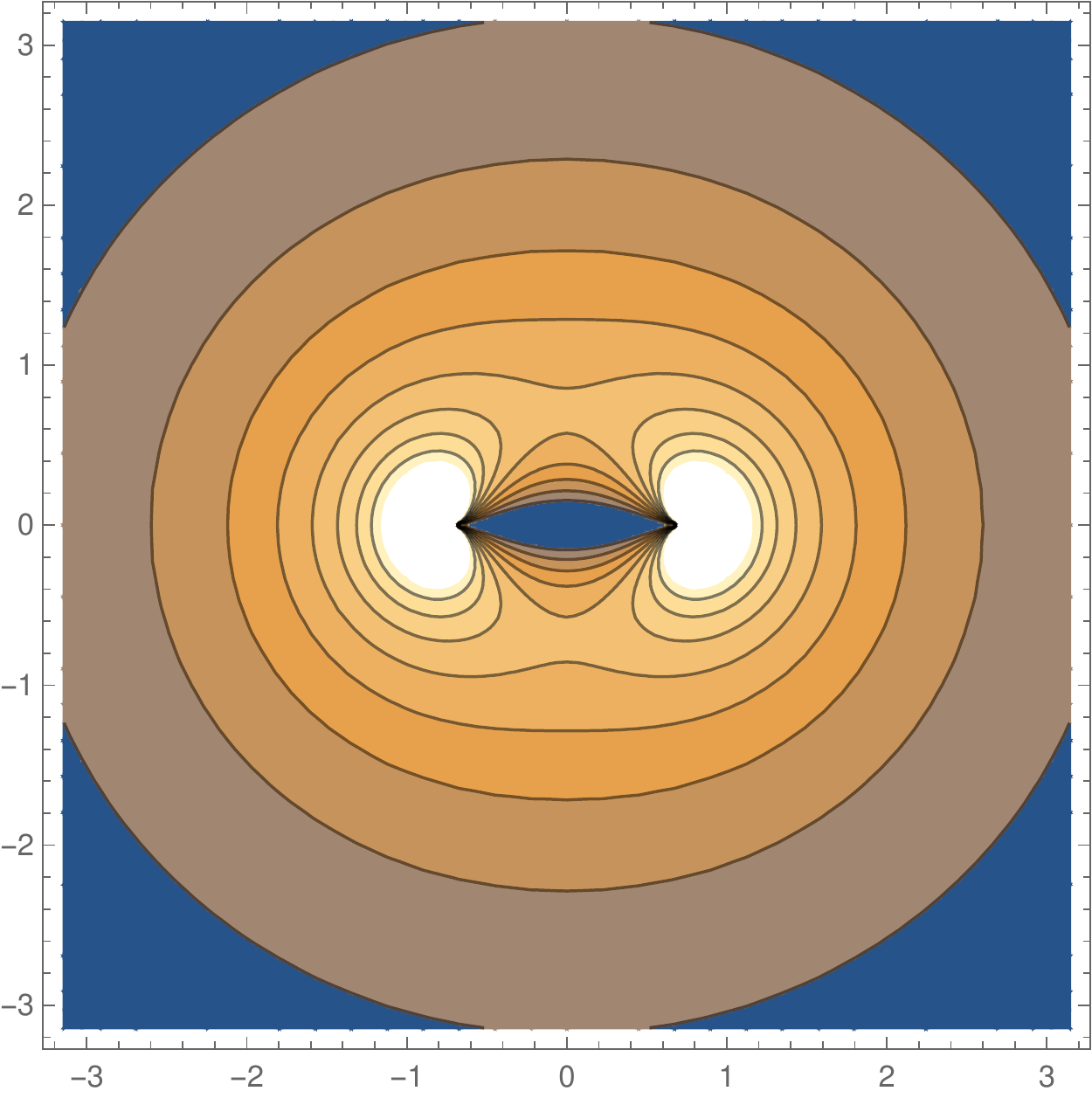}
        \label{fig:potential2}
    \end{minipage}
\caption{Regions of the same colour on the $x-z$-plane (with $z$ pointing upwards) are divided by lines of constant $A_{t}$. The left figure shows the result for $a=0$, the right one for $a=0.7$.}
\label{fig:potential}
\end{figure}

\section{Circular orbits for the $\sqrt{\text{Kerr}}$ potential}
\label{se:AppendixB}
The goal of this section is to recap some basic notions of some linear algebra that are useful to understand the nature of the roots of a third and fourth order degree univariate polynomial. For the polynomial
\begin{align}
\lambda_1 x^{4}+\lambda_2 x^{3}+\lambda_3 x^{2}+\lambda_4 x+\lambda_5=0,
\label{eqn:poly_x}
\end{align}
the explicit roots are given by
\begin{align}
x_{1,2}&=-\frac{\lambda_2}{4 \lambda_1}-S \pm \frac{1}{2} \sqrt{-4 S^{2}-2 p+\frac{q}{S}} \nonumber \\
x_{3,4}&=-\frac{\lambda_2}{4 \lambda_1}+S \pm \frac{1}{2} \sqrt{-4 S^{2}-2 p-\frac{q}{S}},
\end{align}
where we have defined
\begin{align}
p &:=\frac{8 \lambda_1 \lambda_3-3 \lambda_2^{2}}{8 \lambda_1^{2}}  \nonumber \\
q &:=\frac{\lambda_2^{3}-4 \lambda_1 \lambda_2 \lambda_3+8 \lambda_1^{2} \lambda_4}{8 \lambda_1^{3}}  \nonumber \\
\Delta_{0}&:=\lambda_3^{2}-3 \lambda_2 \lambda_4+12 \lambda_1 \lambda_5  \nonumber \\
\Delta_{1}&:=2 \lambda_3^{3}-9 \lambda_2 \lambda_3 \lambda_4+27 \lambda_2^{2} \lambda_5+27 \lambda_1 \lambda_4^{2}-72 \lambda_1 \lambda_3 \lambda_5 \nonumber \\
Q&:=\sqrt[3]{\frac{\Delta_{1}+\sqrt{\Delta_{1}^{2}-4 \Delta_{0}^{3}}}{2}}\nonumber \\
S&:=\frac{1}{2} \sqrt{-\frac{2}{3} p+\frac{1}{3 \lambda_1}\left(Q+\frac{\Delta_{0}}{Q}\right)}.
\end{align}
It turns out that if the discriminant \cite{janson2010roots}
\begin{align}
\Delta := \frac{1}{27} \left(4 \Delta_{0}^3 - \Delta_{1}^2\right) < 0,
\label{eqn:det}
\end{align}
then the equation \eqref{eqn:poly_x} has two distinct real roots and two complex (conjugate) roots. In the particular case $\lambda_1 = 0$, we can use the reduced discriminant $\Delta_R$
\begin{align}
\Delta_R := \lambda_3^2 \lambda_4^2 - 4 \lambda_2 \lambda_4^3 - 4 \lambda_3^3 \lambda_5 + 18 \lambda_2 \lambda_3 \lambda_4 \lambda_5 - 27 \lambda_2^2 \lambda_5^2
\label{eqn:detr}
\end{align}
which is positive when the third order degree polynomial has three real roots and negative when it has one real and two complex conjugate roots. We can apply these tools to find how many real solutions we have for the $x$ variable in the case of circular orbits in  $\sqrt{\text{Kerr}}$: using from the polynomial equation \eqref{eqn:x_eq} where
\begin{align}
\lambda_1 &:= m u \nonumber \\
\lambda_2 &:= 4 a L_{\text{crit}} m u^2 \left(a^2 u^2+1\right) \nonumber \\
\lambda_3 &:= -L_{\text{crit}}^2 m u \left(a^2 u^2+1\right)-2 a^2 m u \nonumber \\
\lambda_4 &:= 4 a L_{\text{crit}} m \left(a^2 u^2+1\right) \nonumber \\
\lambda_5 &:= a^4 m u-\frac{L_{\text{crit}}^2 m \left(a^2 u^2+1\right)}{u}.
\end{align}
we find that
\begin{align}
\Delta = &-\! 16 m^6 \left(a^2 u^2+1\right)^4 \! \left(2 a^2 L_{\text{crit}} u^2 \left(2 a^2 \!+\! L_{\text{crit}}^2\right) \!+\! L_{\text{crit}}^3\right)^2 \nonumber \\
&\times \Big\{L_{\text{crit}}^2 u^2 \left[27 a^4 u^4 + L_{\text{crit}}^2 a^2 u^4 \right. \nonumber \\
& \hspace{70pt} \left. +36 a^2 u^2 +L_{\text{crit}}^2 u^2 +8\right]+16 \Big\}
\end{align}
is always manifestly negative.

\bibliographystyle{apsrev4-2}
\bibliography{inspire}
\end{document}